\documentclass[aps,amsmath,amssymb,groupedaddress,twocolumn]{revtex4}
\usepackage{graphicx}

\newcommand{\nc}{\newcommand}
\nc{\bs}{\bigskip}
\nc{\beq}{\begin{equation}}
\nc{\eeq}{\end{equation}}
\nc{\beqa}{\begin{eqnarray}}
\nc{\eeqa}{\end{eqnarray}}

\def\gsim{\mathrel{\rlap{\lower4pt\hbox{\hskip1pt$\sim$}}
    \raise1pt\hbox{$>$}}}       

\begin{document}

\title{White holes and eternal black holes}

\author{Stephen~D.~H.~Hsu} \email{hsu@uoregon.edu}\affiliation{Institute of Theoretical Science, University of Oregon, Eugene, OR 97403 }

\date{November 2011}

\begin{abstract}
We investigate isolated white holes surrounded by vacuum, which correspond to the time reversal of eternal black holes that do not evaporate. We show that isolated white holes produce quasi-thermal Hawking radiation. The time reversal of this radiation, incident on a black hole precursor, constitutes a special preparation that will cause the black hole to become eternal.
\end{abstract}


\maketitle


\bigskip

{\bf What is a white hole?} \bigskip

White holes have received far less attention from researchers than black holes (for a review, see, e.g., \cite{FN}). This is understandable, given that conditions in our universe readily lead to black hole formation, whereas white hole creation has neither been observed nor is expected to have occurred in the history of the universe. 

However, white holes are themselves fundamental objects and worthy of further study. White holes are time-reversed black holes, and therefore characterized by the same quantum numbers: mass, angular momentum, charge. While a classical black hole spacetime has a singularity in the future, a white hole has one in the past. If quantum gravitational effects can resolve black hole singularities, then white holes need not result from singular initial conditions. (In any case the initial white hole singularity is not directly visible to observers.)

Standard quantum mechanical reasoning suggests that any initial state which evolves into a black hole also has some nonzero probability to evolve into a white hole.  Note we are referring to a  quantum gravitational process, and are making the assumption that even in quantum gravity tunneling between two states with the same quantum numbers has non-zero (although possibly very small) probability. For example, it is known that the collision of two sufficiently energetic particles can create a black hole \cite{bh}. Because the quantum numbers are the same we would expect that the same energetic particles have a small but non-zero probability of producing a white hole with the same quantum numbers \cite{quantum}. Since large black holes are long-lived, there are some white hole states (corresponding to time slices late in the black hole's existence) that persist for a long time before exploding. Thus, long-lived white holes are a consequence of quantum mechanics and the properties of black holes. 

A class of highly entropic objects whose full spacetime evolution is that of a white hole which explodes outwards, is stopped by gravitational self-attraction, and recollapses to form a black hole, are described in \cite{monsters}.

\bigskip
{\bf Hawking's arguments and thermal equilibrium}
\bigskip

In his 1976 paper {\it Black holes and thermodynamics} \cite{HBT}, Hawking  analyzed the properties of white holes by considering a box in thermal equilibrium, whose temperature and volume are adjusted so that the most probable configuration is a black hole surrounded by a gas of particles whose temperature is equal to that of a black hole. The black hole emits Hawking radiation but absorbs, on average, as much energy from the gas as it emits. Applying time reversal, the configuration describes a white hole emitting and absorbing radiation. Since there is no arrow of time for a system in thermal equilibrium, Hawking argued that black and white holes must be indistinguishable. More precisely, the properties of white and black holes {\it in equilibrium with their surroundings} are identical. However, the same cannot be said for black and white holes in isolation (i.e., surrounded by empty vacuum)-- we shall see that their properties are radically different.

In elementary particle physics we are accustomed to the idea that time reversal maps particles to their anti-particles. However, in the case of black and white holes the subsequent evolution of the time-reversed object depends on more than just quantum numbers such as $M, J, Q$. For a hot black hole in a cold environment, there is a statistical arrow of time.

\bigskip
{\bf Isolated white holes}
\bigskip

\begin{figure}[ht]
\includegraphics[width=8cm]{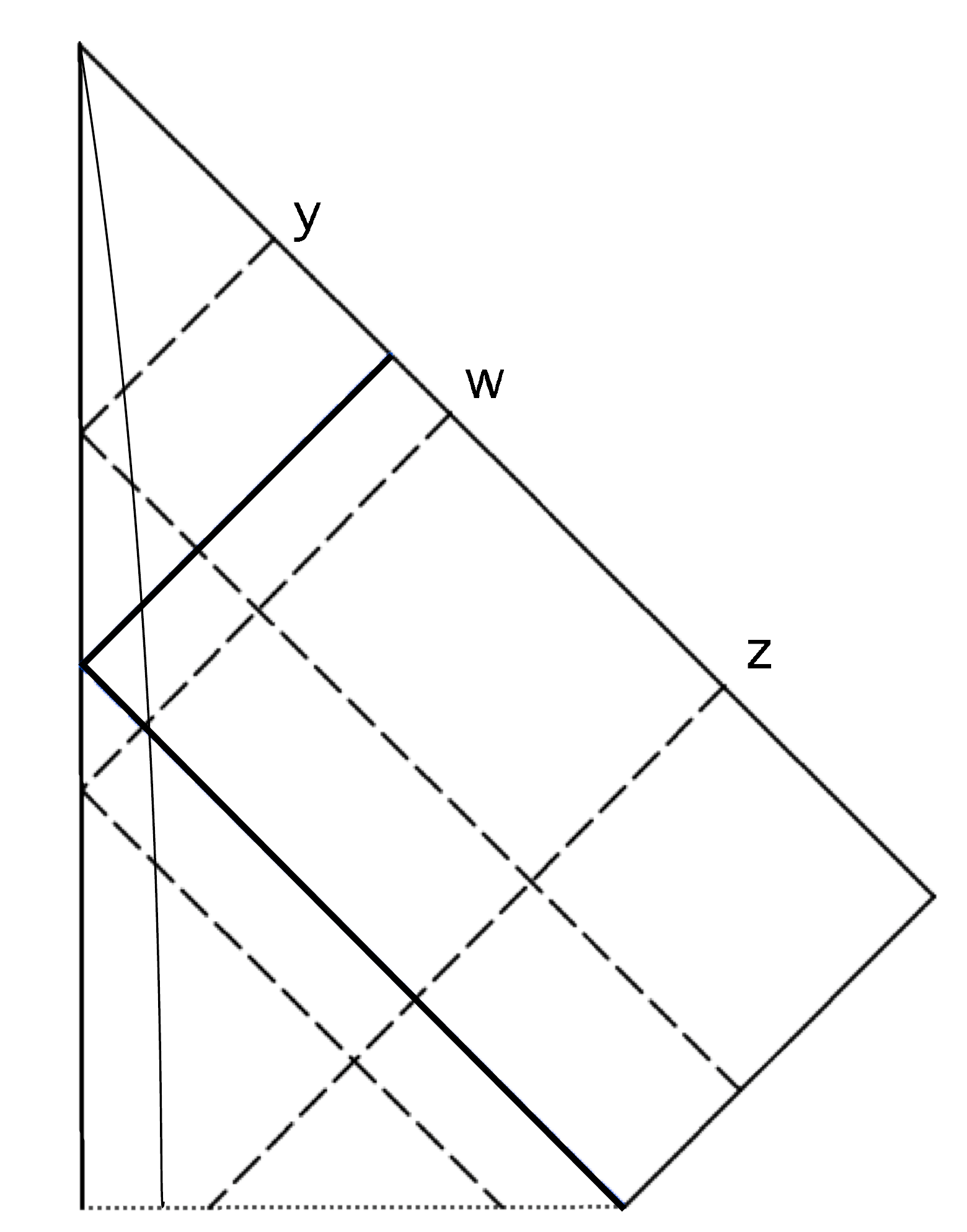}
\caption{A white hole spacetime. We impose the condition that past null infinity $\cal{J}_-^{\rm wh}$ is in the vacuum state -- there is no incoming radiation from the far past. The dotted black line is the initial singularity, and the thick solid line is the path of a null ray on the anti-horizon. The curved line indicates matter which explodes out of the hole. The dashed black lines refer to modes discussed in the text.}
\label{figure1}
\end{figure}

\begin{figure}[ht]
\includegraphics[width=8cm]{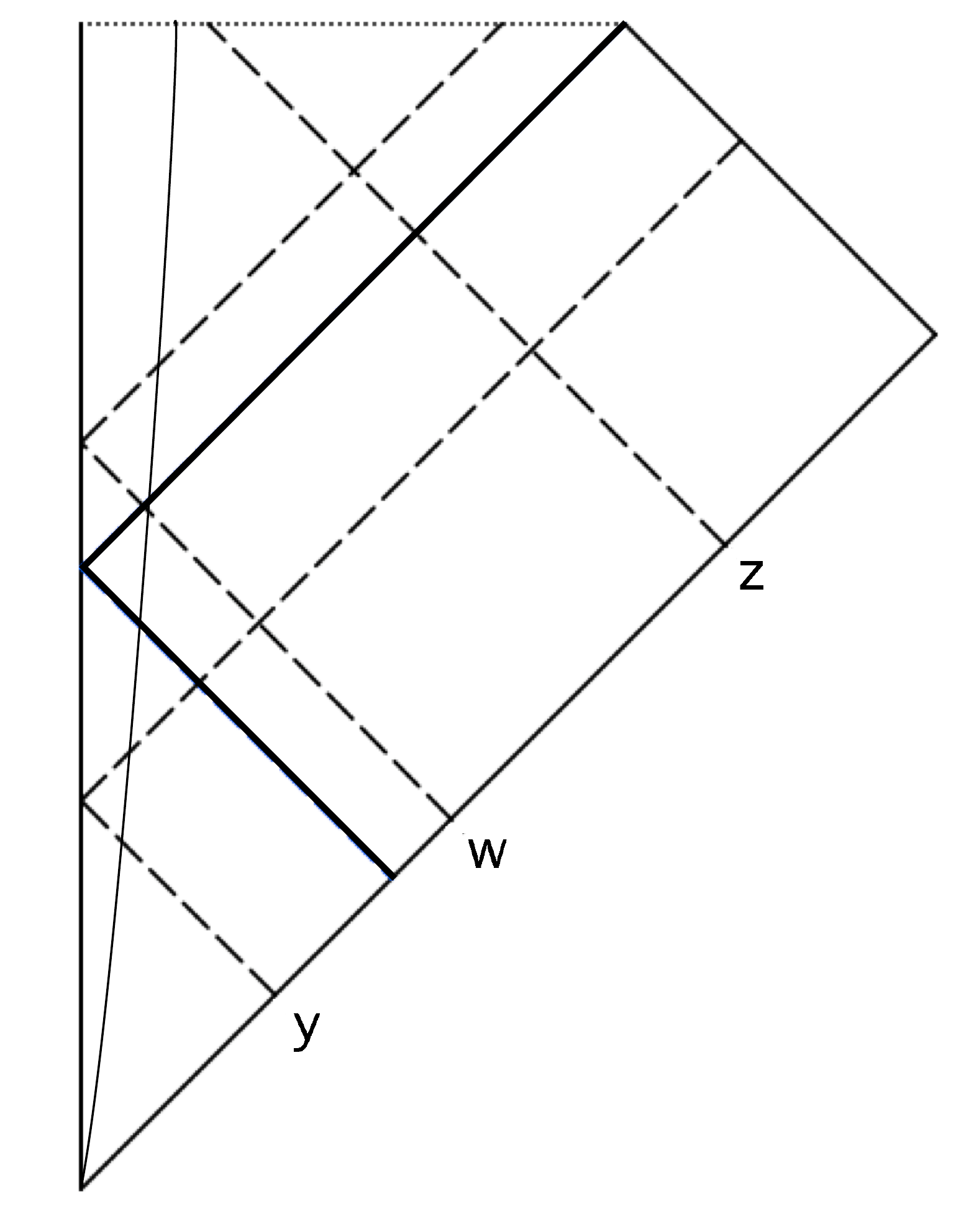}
\caption{A black hole spacetime which is the time reversal of figure 1. We impose the condition that future null infinity $\cal{J}_+^{\rm bh}$ is in the vacuum state -- there is no outgoing Hawking radiation. The dotted black line is the final singularity, and the thick solid line is the path of a null ray which coincides with the horizon at late times. The curved line indicates matter which collapses to form the hole. The dashed black lines refer to modes discussed in the text. 
}
\label{figure2}
\end{figure}

Consider a white hole (figure 1) in a spacetime with the property that past null infinity is in the empty vacuum state of ordinary flat space. This implies that space far from the hole is empty and that there is no incoming radiation from the past. This white hole spacetime is the time reversal of a black hole spacetime with no Hawking radiation propagating to future infinity (figure 2). One motivation for considering such objects is that they are localized, as opposed to an entire spacelike slice of a box in thermal equilibrium. Do such white holes exist? What are their properties? (For simplicity, we assume all quantum numbers of the hole, other than its mass, are zero.)

In this discussion we will refer to the diagrams in figures 1 and 2, which depict a black hole spacetime (figure 2) and its time reversal (figure 1). We will refer to past and future null infinity of the black hole spacetime as $\cal{J}_\pm^{\rm bh}$, where the subscript $+$ indicates future, and $-$ indicates past. In the white hole diagram the role of past and future are reversed: $\cal{J}_\mp^{\rm wh} = \cal{J}_\pm^{\rm bh}$.

\smallskip
Assuming that the white hole is isolated implies the empty vacuum on $\cal{J}_-^{\rm wh}$ : there is no incoming radiation from the past.

This condition is equivalent, on the black hole spacetime, to no Hawking radiation propagating to future null infinity $\cal{J}_+^{\rm bh}$. This sounds strange, but can be accomplished by proper choice of initial state from which the black hole is formed. That is, a special arrangement of incoming modes from $\cal{J}_-^{\rm bh}$ is required; see below for details. In the white hole spacetime these modes would be seen exiting the white hole after it explodes from behind its anti-horizon.

In our discussion we {\it assume} that the black hole spacetime (figure 2) describes a progenitor (e.g., a star) which collapses to form the hole. Because ordinary stars and other progenitors in nearly-flat space obey an entropy bound: $S < A^{3/4}$, where $A$ is their surface area in Planck units, such objects have much lower entropy than a black hole with no constraint on how it was formed \cite{Frampton:2008mw,monsters}. Indeed, a generic black hole, formed in a maximally entropic process (e.g., by allowing an initially small black hole to slowly accrete matter) has entropy of order $A$, but such objects do not satisfy the isolation condition imposed above. Thus, the objects under study are very exotic (improbable): not only are they white holes, but the condition of isolation further reduces the entropy substantially. Our analysis is mainly of theoretical, rather than practical astrophysical, interest.

In our analysis we only treat the case of a Schwarzschild black hole. Holes with angular momentum or electric charge have a more complex inner structure, including a Cauchy horizon. Interactions between outgoing and backscattered ingoing radiation near this Cauchy horizon lead to a curvature singularity known as mass inflation \cite{MI}. The resulting inner structure seems to involve quantum gravitational effects and is still not completely understood.

Following Hawking \cite{H1976}, we define an orthogonal set of modes for a scalar field on the black hole spacetime.
\begin{equation}
\phi =  \int d\omega  \left( f_\omega a_\omega  + \bar{f}_\omega a^\dagger_\omega \right)~~,
\end{equation}
where the $\{ f_\omega \}$ are a complete, orthonormal family of complex solutions of the wave equation. The notation used here is identical to that in \cite{H1976}, except that (see below) we define the destruction operators of the $w,y,z$ modes to be $a^w, a^y, a^z$ rather than $g,h,j$ as Hawking did. In all other respects, we adhere to his definitions, which we now briefly review. 

The modes are defined on an extended Schwarzschild geometry, obtained by analytic continuation, which includes both past and future horizons, ${\cal H}^\pm$, but does not describe the collapse which forms the black hole. Let $u$ and $v$ be the retarded and advanced time coordinates for the Schwarzschild metric. Let $U$ (Kruskal coordinate) be the affine parameter on the the past horizon ${\cal H}^-$:
\beq
u = - 4M \ln ( -U )~~,
\eeq
where $M$ is the mass of the black hole. The $\{ f^{(1)}_\omega \}$ modes are solutions on the extended spacetime with zero Cauchy data on the past horizon and time dependence of the form $\exp (i \omega v)$ on $\cal{J}_-^{\rm bh}$. The $\{ f^{(2)}_\omega \}$ are solutions with zero Cauchy data on $\cal{J}_-^{\rm bh}$ and dependence $\exp (i \omega U)$ on the past horizon. The analytic continuation of $u$ yields two coordinates 
\beq
u_\pm = - 4M ( \ln U \ \mp \ i \pi )~~~~ (U > 0)~,
\eeq
with $u_+ = u_-$ for $U<0$. These are used to replace the $f^{(2)}$ modes by two orthogonal families of solutions $f^{(3)}$ and $f^{(4)}$. These have zero Cauchy data on $\cal{J}_-^{\rm bh}$ and dependence on the past horizon of the form $\exp (i \omega u_+)$ and $\exp (i \omega u_-)$, respectively.

The physical interpretation of these modes, after continuation back to the collapse spacetime, is as follows. The $f^{(1)}$ modes enter the black hole after a horizon has formed. The $f^{(2)}$ modes (equivalently, the $f^{(3),(4)}$ modes) enter the black hole region before the $f^{(1)}$ modes, at earlier advanced time; in the time-reversed spacetime they would exit the white hole before it emerges from behind its anti-horizon (see figure 1). The quantum states associated with these modes are observable to a detector at $\cal{J}_-^{\rm bh}$, or equivalently at $\cal{J}_+^{\rm wh}$.  We define the destruction operators for the $f$ modes to be $a^{1}, a^{3}, a^{4}$.

It is useful to define additional bases of modes (see figures): $\{ w_\omega \}$, $\{ y_\omega \}$ and $\{ z_\omega \}$, which are linear combinations of the $\{ f^{(i)}_\omega \}$, and are observable by a detector at $\cal{J}_+^{\rm bh}$. The $\{ w_\omega \}$ modes have zero Cauchy data on $\cal{J}_-^{\rm bh}$ and on the past horizon for $U < 0$. For $U > 0$ on the past horizon their dependence is of the form $\exp (-i \omega u_+)$. The $\{ y_\omega \}$ modes have zero Cauchy data on $\cal{J}_-^{\rm bh}$ and on the past horizon for $U > 0$. For $U < 0$ on the past horizon their dependence is of the form $\exp (i \omega u_+)$. The $\{ z_\omega \}$ modes are identical to the $f^{(1)}$ modes already defined. The destruction operators for these new modes are $a^w, a^y, a^z$.

The physical interpretation of these modes, after continuation back to the collapse spacetime, is as follows. The $y$ modes enter the spatial region where the black hole will be formed (i.e., the precursor), but emerge before a horizon appears. The transmitted $y$ modes (which are not reflected by the gravitational potential back into the hole) are observable at future null infinity of the black hole spacetime, $\cal{J}_+^{\rm bh}$. The $w$ modes propagate in from $\cal{J}_-^{\rm bh}$, enter the black hole region of space (i.e., the precursor) before a horizon is formed, but are trapped and encounter the future singularity. In the white hole spacetime (see figure 1), $w$ modes emerge first from the anti-horizon, followed by the $y$ modes, which appear after matter begins to explode from behind the anti-horizon. 

The Hawking radiation modes $p_\omega$, which are observable by a detector at $\cal{J}_+^{\rm bh}$, are a complete set of orthonormal solutions which contain only positive frequencies at $\cal{J}_+^{\rm bh}$ and are purely outgoing (zero Cauchy data on the horizon of the collapse spacetime). They can be written in terms of the $y$ and $z$ modes \cite{H1976}:
\beq
\label{pmode}
p_\omega = t_\omega y_\omega + r_\omega z_\omega~~,
\eeq
and the destruction operator for this mode is 
\beq 
a^p_\omega = \bar{t}_\omega a^y_\omega + \bar{r}_\omega a^z_\omega~~.
\eeq
Equation (\ref{pmode}) can be understood as follows from figure 2, depicting the black hole spacetime. The modes which reach future infinity are a superposition of transmitted $y$ modes and reflected $z$ modes, where $t$ and $r$ are the transmission and reflection amplitudes for waves incident on the black hole.

The condition that $\cal{J}_+^{\rm bh}$ is in the vacuum state (no Hawking radiation; an {\it eternal} black hole) is
\beq
\label{empty}
a^p_\omega \, \vert 0_+^{\rm bh} \rangle = 0~.
\eeq
This condition is not typically imposed on the future state of the black hole spacetime. Instead, one usually requires that the precursor state (i.e., a collapsing star) is surrounded by vacuum, which is a condition on the past rather than on the future. However, the time reversal symmetry of quantum field theory and of general relativity imply that there must exist initial conditions that lead to the future condition (\ref{empty}). In the white hole case it is natural to impose the vacuum condition on the past, and we explore what its consequences are for the future of the hole.

A sufficient, but not necessary, condition for satisfying (\ref{empty}) is to require
\beq
\label{sufficient}
a^y_\omega \, \vert 0_+^{\rm bh} \rangle =  a^z_\omega \, \vert 0_+^{\rm bh} \rangle = 0~.
\eeq
In his original discussion of the future vacuum on the black hole spacetime \cite{H1976}, Hawking imposes (\ref{sufficient}) as well as the additional condition 
\beq
\label{special}
a^w_\omega \, \vert 0_+^{\rm bh} \rangle = 0 ~,
\eeq 
requiring that the future vacuum be empty of unnecessary $w$ modes. In our discussion of white holes this condition need not apply since we do not wish to constrain the initial state of the white hole other than to require its isolation.

It is straightforward to calculate the particle number content, mode by mode, for the state defined above. We are specifically interested in the $f^{(i)}$ modes, which are detectable as particles incident on the black hole and its precursor by an observer at $\cal{J}_-^{\rm bh}$ (past infinity of the black hole spacetime). Equivalently, these modes are detectable as outgoing particles by an observer far outside the white hole. The condition $a^z_\omega \, \vert 0_+^{\rm bh} \rangle = 0$, imposed in (\ref{sufficient}), is identical to the condition $a^1 \, \vert 0_+^{\rm bh} \rangle = 0$, which implies that there are no $f^{(1)}$ or $z$ modes emitted by the white hole (or absorbed by the eternal black hole). These modes are emitted by the white hole long before it explodes from behind its anti-horizon, or equivalently are absorbed by the black hole long after its horizon forms (see figures). The remaining $f^{(3,4)}$ modes are linear combinations of the $w$ and $y$ modes, which are emitted by the white hole just before and after it explodes. The $y$ modes, in particular, appear to be emitted from the ejecta of the hole.

We obtain
\beq
\langle 0_+^{\rm bh} \vert  \, a^{3 \, \dagger}_\omega a^3_\omega + a^{4 \, \dagger}_\omega a^4_\omega \,\vert 0_+^{\rm bh} \rangle = \frac{2x}{1-x} + \frac{1+x}{1-x} \langle 0_+^{\rm bh} \vert \, a^{w \,\dagger}_\omega a^w_\omega \,\vert 0_+^{\rm bh} \rangle~,
\eeq
where $x = \exp (- \beta \omega )$ and $\beta$ the Hawking temperature of the black hole. In the simple case with $a^w_\omega \vert 0_+^{\rm bh} \rangle = 0$, the particle occupation numbers of each of the $f^{(3)}$ and $f^{(4)}$ modes are simply those of the blackbody distribution ($i = 3$ or $4$): 
\beq
\langle 0_+^{\rm bh} \vert  \, a^{i \, \dagger}_\omega a^i_\omega \vert 0_+^{\rm bh} \rangle = \frac{x} {1-x} = \frac{1}{\exp (\beta \omega) - 1}~~~.
\eeq 
Physically, this means that one can construct an {\it eternal} (i.e., non-radiating) black hole in the minimal state satisfying (\ref{sufficient}) and (\ref{special}) by exposing its precursor (and, briefly, the black hole itself) to a special quasi-thermal radiation state. It also implies that an isolated white hole in the state satisfying (\ref{special}) will radiate quasi-thermally just before and after it explodes from behind its anti-horizon. Note that although the occupation numbers we have calculated are thermal, the state is actually a pure state if the initial white hole state was pure. Unlike in the case of Hawking radiation, we are not forced to trace over any causally disconnected region, and we do not necessarily obtain a mixed state description. 

In our analysis so far we have treated the background spacetime as fixed and have neglected backreaction effects. In the original Hawking analysis, one first obtains the thermal spectrum of black hole radiation, and then invokes energy conservation and backreaction to argue that the hole steadily loses mass through radiation, eventually (perhaps) evaporating completely. Since the rate of energy loss is small it is assumed that the semiclassical analysis pertains until the final Planckian stage of evaporation. In our case we can make the same argument regarding the white hole: our calculations initially assume a fixed spacetime, but lead to thermal behavior of the hole just before and after it explodes. Conservation of energy implies that the white hole and the ejecta somehow compensate for the emitted radiation so that the total energy that reaches infinity is the initial ADM mass of the hole. How this happens is not entirely clear, although one can simply regard it as a constraint on possible final states resulting from an isolated white hole. We note that the mode bases used in this analysis only depend on the asymptotic structure of the black or white hole spacetime. The details of how the black hole is formed, or how the white hole explodes, do not affect the results; indeed, the analysis can be formulated on the extended Schwarzschild spacetime  which is the analytic continuation of the realistic geometry which contains a collapsing/exploding body.

\bigskip

A necessary and sufficient condition for isolation of the white hole (as opposed to (\ref{sufficient}), which was sufficient, but a special case) is 
\beq
\label{necessary}
\bar{t}_\omega a^y_\omega \, \vert 0_+^{\rm bh} \rangle = -  \bar{r}_\omega a^z_\omega \, \vert 0_+^{\rm bh} \rangle~.
\eeq
As mentioned previously, the condition (\ref{necessary}) can be understood (see figure 2) as the requirement that reflected $z$ modes interfere perfectly with transmitted $y$ modes so that no Hawking radiation reaches future infinity of the black hole spacetime. This, more general, condition allows for Hawking-like radiation from the white hole in the form of $z$ modes, which leave the white hole long before its explosion and reach future infinity  $\cal{J}_+^{\rm wh}$.

In the general case, we obtain the following expression for $f^{(3,4)}$ mode occupation numbers:
\beqa
&& \langle 0_+^{\rm bh} \vert  \, \sum_{i=3,4} \, a^{i \, \dagger}_\omega a^i_\omega \, \vert 0_+^{\rm bh} \rangle = \frac{1}{1-x} \Big[ 2x  +    \langle 0_+^{\rm bh} \vert \,  \nonumber (1+x) \times  \, \\ &&  ( a^{y \, \dagger} a^y + a^{w \,\dagger} a^w)  - 2\sqrt{x} (a^w a^y  +   a^{w \, \dagger} a^{y \, \dagger}            
) \,    \vert 0_+^{\rm bh} \rangle  \Big]~.
\eeqa
For $\vert 0_+^{\rm bh} \rangle$ which are particle number eigenstates of $y$ and $w$, 
one can use the Cauchy-Schwarz inequality $$\vert \langle a^w a^y \rangle \vert^2 \leq \langle \, a^{w \, \dagger} a^w ~ a^{y \,\dagger} a^y \, \rangle~,$$ and identities $(1+x) \geq 2 \sqrt{x}$ and $N_y + N_w \geq 2 \sqrt{N_w N_y}$, to see that the expectation value of the mode number is, for every frequency, at least as large as in the simplest case where the conditions (\ref{sufficient}) and (\ref{special}) are satisfied.

For a white hole to be indistinguishable from an ordinary black hole it must emit Hawking-like radiation from the beginning, with thermal occupation numbers for $\langle a^{z \,\dagger} a^z \rangle$.  Condition (\ref{necessary}) then requires non-zero occupation numbers for $\langle a^{y \, \dagger} a^y \rangle$, leading to more energy radiated in $f^{(3,4)}$ modes at late times. A significant amount of energy in this form must emerge {\it after} the white hole explodes, which limits how much can be radiated before it explodes. It is hard to see how an isolated white hole can behave so as to be indistinguishable from an ordinary black hole of equal mass. This only seems possible if we remove the condition of isolation, allowing the white hole to both emit and absorb energy, as would be the case for the thermal box considered originally by Hawking \cite{HBT}.

\bigskip
{\bf Conclusions}
\bigskip

We summarize our main results below. These results have not, to our knowledge appeared previously in the literature.

1. Isolated white holes behave very differently from isolated black holes. This is due to the lack of time reversal symmetry in the surrounding environment: the statistical arrow of time implies that isolated black holes evaporate into their cold surroundings, whereas isolated white holes are, by definition, not bathed in incident radiation. Complete time reversal symmetry is only present in thermal equilibrium, the case originally analyzed by Hawking. 

2. Isolated white holes with initial state given by the simple conditions (\ref{sufficient}) and (\ref{special}) emit quasi-thermal radiation just before and after exploding from behind their anti-horizon. Modifying the initial state, while retaining the condition of isolation, likely implies even more radiation at late stages. There do not seem to be isolated white holes which are indistinguishable from isolated black holes of the same mass. 

3. As a byproduct of our investigation, we note the existence of eternal -- non-evaporating -- black holes, formed from special quantum initial states. We do not know whether such holes are stable against perturbations. That is, if one prepares a black hole in this ``eternal'' state, but the hole subsequently interacts with a small probe (whose existence was not anticipated in the original preparation), does this cause only a small leakage of Hawking radiation, or does the hole revert to ordinary evaporation? Another interesting question is the relative entropy of eternal and ordinary black holes.


\smallskip

\emph{Acknowledgments ---} The author thanks Roberto Casadio, Ted Jacobson and David Reeb for comments or discussions. This work is supported by the Department of Energy under grant DE-FG02-96ER40969.

\bigskip



\begin{thebibliography}{99}
\bibitem{quantum}
To be more specific, take a spacelike slice (e.g., horizontal line) near the top of figure 2 (black hole Penrose diagram), but below the singularity. The time reversal of this state has the same values (and time derivatives) of metric and matter fields as a corresponding spacelike slice near the bottom of figure 1 (white hole diagram). If colliding particles can tunnel to the first (black hole) state, they can also tunnel to the corresponding white hole state (i.e., time reversal does not alter any quantum numbers). Note that because this is a quantum process, there is no singularity in the past of the white hole created. The spacetime with two colliding particles is simply patched onto the initial condition described by the spacelike slice from figure 1.
\end{thebibliography}
\end{document}